\documentclass[galaxies,review,accept,oneauthor,pdftex,12pt,a4paper]{mdpi} 
\lastpage{x}
\doinum{10.3390/------}
\pubvolume{xx}
\pubyear{2014}
\history{Received: xx / Accepted: xx / Published: xx}
\pdfoutput=1

\usepackage{graphicx}

\Title{The Third Law of Galactic Rotation}

\Author{Stacy S.\ McGaugh $^{1}$}

\address{%
$^{1}$ Department of Astronomy, Case Western Reserve University, Cleveland, OH 44118, USA}

\corres{stacy.mcgaugh@case.edu, phone: +1-216-368-1808, fax: +1-216-368-5406.}

\abstract{I review the connection between dynamics and the baryonic mass distribution in rotationally supported galaxies.  
The enclosed dynamical mass-to-light ratio increases with decreasing galaxy luminosity
and surface brightness.  The correlation with surface brightness appears to be the more fundamental, with the dependence on luminosity 
following simply from the weaker correlation between luminosity and surface brightness.  In addition to this global relation, 
there is also a local relation between the amplitude of the mass discrepancy and the acceleration predicted by the
observed distribution of baryons.  I provide an empirical calibration of this mass discrepancy--acceleration relation. 
The data are consistent with the operation of a singe effective force law
in disk galaxies, making this relation tantamount to a natural law.
I further provide formulae by which the radial dark matter distribution can be estimated from surface photometry.
The form of the dark matter halo depends uniquely on the distribution of baryons in each galaxy, and in general is
neither a cusp nor a core.  It remains difficult to see how galaxy formation models can reproduce the observed
behavior, which is uniquely predicted by MOND.  
}


\begin{document}


\section{Introduction}

The dynamics of galaxies clearly evince a mass discrepancy \cite{sandersbook}. 
The observed motions of stars and gas in galaxies
cannot be explained by established dynamical laws relating the force of gravity to the observed distribution of
luminous matter.  Nevertheless, the data show a remarkably tight empirical correlation between the observed mass 
distribution and the corresponding dynamics \cite{SandersMDacc,renzorule,MDacc,Swaters2012}.

The properties of rotationally supported galaxies can be encapsulated in three empirical rules.
1.\ Rotation curves attain an approximately constant velocity that persists indefinitely (flat rotation curves) \cite{vera,bosma}.
2.\ The observed mass scales as the fourth power of the amplitude of the flat rotation (the Baryonic Tully-Fisher Relation) 
\cite{TForig,btforig}.
3.\ There is a one-to-one correspondence between the radial force and the observed distribution of baryonic matter
(the mass discrepancy--acceleration relation) \cite{SandersMDacc,MDacc}.

These are statements of the empirically observed behavior of galaxies and their rotation curves.
It is an important question whether these are simply scaling relations that emerge from the process of galaxy formation
in the context of the dark matter paradigm, or genuine laws of nature that follow from a modification of known dynamical laws. 
Both sides of this issue are discussed by \cite{myCJP}.
Here I review the third law, which is less widely appreciated than the first two.

\section{Galaxy Rotation Curves}
\label{sec:RCs}

Galaxies come in two basic dynamical states, disks and spheroids \cite{BM}.
Disks, e.g., spiral and most irregular galaxies, are thin, rotationally supported systems of stars and gas.
Spheroids, e.g., giant ellipticals and dwarf spheroidal galaxies, are pressure supported systems composed mainly of stars.

Dynamically, disks are cold, with rotation velocities well in excess of the local velocity dispersion of stars: 
$V/\sigma \gg 1$ \cite{BT}.  In contrast, spheroids are predominantly pressure supported systems that are dynamically hot, with $V/\sigma < 1$.
The anisotropy of orbits in spheroidal galaxies makes the reconstruction of orbits and inference of the corresponding
gravitational potential challenging \cite{norot,angusdw,serradw}.  In contrast, the thin, dynamically cold nature of galactic disks makes inference of
the radial force relatively straight forward.  Corrections for non-circular motions are generally small 
\cite[if not always negligible][]{OhThings,trachTHINGS,KdN09,WLM} so that the observed rotation curve $V(R)$ is a good tracer of the gravitational potential:

\begin{equation}
- \frac{\partial \Phi}{\partial R} = \frac{V^2}{R}.
\label{eqn:potent}
\end{equation}

\begin{figure}
\includegraphics[width=17cm]{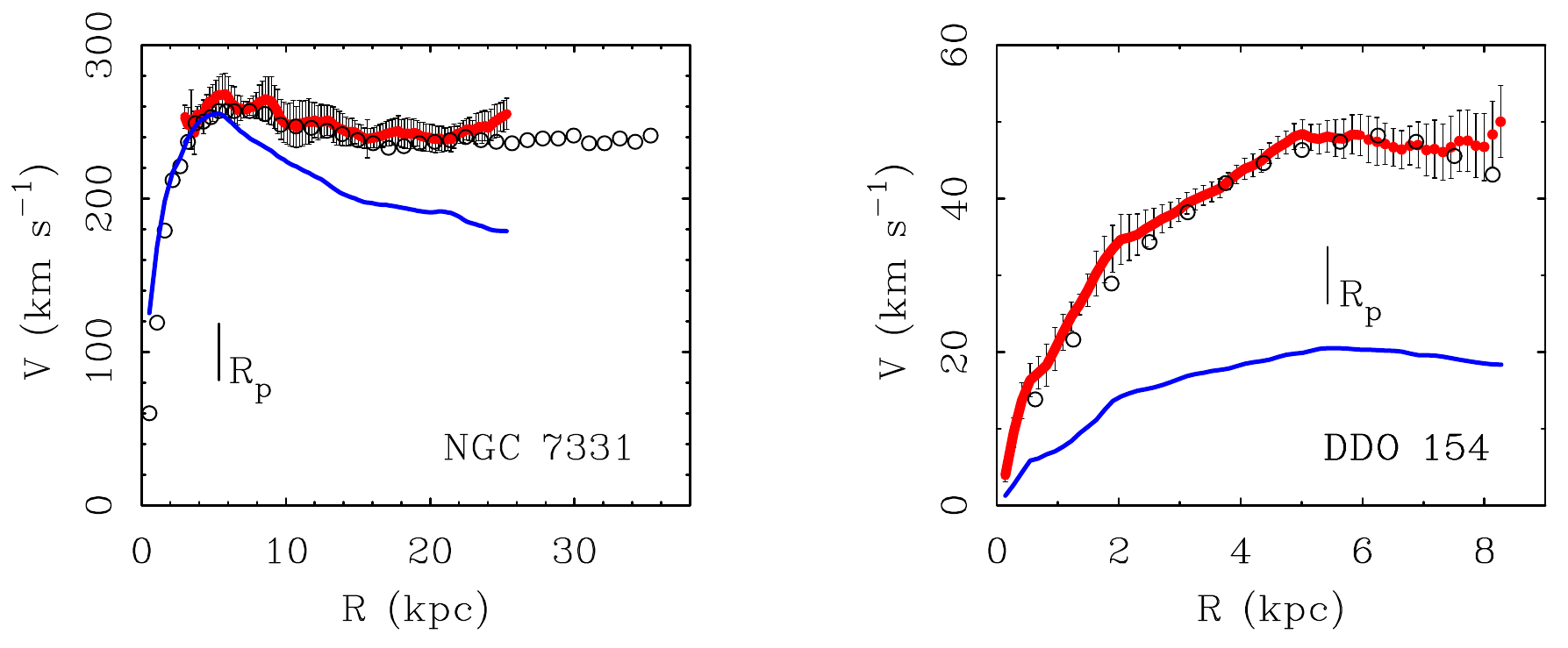}
\caption{The rotation curves of NGC 7331 (left) and DDO 154 (right).  Note the difference in scale.
NGC 7331 is a massive spiral galaxy with a strong central concentration of stars.  DDO 154 is a
diffuse, gas rich, low surface brightness dwarf galaxy.  Data from the THINGS survey
\cite[red points with error bars]{THINGS} are shown together with older data \cite[open points]{BBS,CPddo154}.
Blue lines show the baryonic mass model, which is the
rotation velocity attributable to the observed stars and gas, $V_b(R)$.  The radius where $V_b$ peaks is marked $R_p$. 
The excess rotation above the line representing the baryons illustrates the mass discrepancy in galaxies.
}
\label{fig:RC}
\end{figure}

Fig.~\ref{fig:RC} shows the rotation curves of two galaxies that illustrate the behavior of disk galaxies of very different
characteristics.  In both cases, the rotation curves become approximately flat at large radii.  This contrasts with the
expected Keplerian decline once the majority of stars and gas are enclosed, providing one of the clearest examples of
the mass discrepancy.  

The flatness of rotation curves presumably cannot persist indefinitely.
The implied mass diverges: since $M \propto V^2 R$ and $V \approx$ constant, the enclosed dynamical mass
increases linearly with radius.
Nevertheless, the observer's experience is that this behavior persists as far out as reliable tracers can be observed.
The occasional hint of a downturn in the outer parts\footnote{I refer here to Keplerian downturns
at large radii.  These should not be confused with rotation curves that fall somewhat at small radii before flattening out,
a common morphology in bright galaxies.} of rotation curves has yet to be confirmed.
For example, it was suggested that the total mass of DDO 154 had been encompassed \cite{CPddo154}, but 
this was not confirmed by subsequent observation \cite{THINGS}.  So far, this has been the general experience \cite{FTHINGS}.

There are indications from the motions of satellite galaxies and gravitational lensing that the implied total mass of
galaxies is finite \cite[e.g.,][]{ZW94,Prada03,SDSSlens,Reyes}.
These are statistical measures that apply in an average sense across many galaxies rather than to individual objects.
The constraint extends to much larger radii than provided by observations of rotation curves, but lacks their
specificity and precision: one does not see the orbits of individual objects within specific galaxies as one does with 
rotation curves.  Broadly speaking, the results from statistical methods are consistent with rotation curves that 
remain approximately flat or slowly decline until galaxies bump into one another \cite{MdB98a}.  
That is to say, the apparent edge of galaxy mass distributions implied by the various
statistical measures corresponds to the radius where the influence of surrounding galaxies becomes comparable
to that of the galaxy under consideration.

In addition to the tendency towards asymptotic flatness, the galaxies in Fig.~\ref{fig:RC} illustrate a number of
other generic aspects of observed rotation curves.  For example, the amplitude of the flat velocity is a strong function
of the mass of stars and gas: $M_b \propto V_f^4$. This Baryonic Tully-Fisher relation \cite{btforig,M05,M12} allows one
to estimate the relative masses of rotating galaxies.  NGC 7331 rotates faster than DDO 154, so it is more
massive --- by about a factor of 500. 

The flat rotation velocity scales with the total baryonic mass, without reference to dark matter.  
It does not matter whether the baryonic mass\footnote{That the Baryonic Tully-Fisher relation is very tight once both the observed
stars and gas are included implies that there are no large reservoirs of unseen baryons in disk galaxies \cite{btforig}.} 
is in stars or gas.  It is insensitive to the distribution of this mass.
Galaxies of the same baryonic mass have the same $V_f$ irrespective of whether the stars are concentrated or spread thin \cite{zwaanTF,MdB98a,CR,myPRL}.

In contrast, the shape of the inner rotation curve is quite sensitive to the distribution of baryons \cite{sellwood1999,PP00,noordURC,Lelli13,Lelli14}.
Galaxies with a strong central concentration of stars have rapidly rising rotation curves, while diffuse galaxies
have slowly rising rotation curves (Fig.~\ref{fig:RC}).  The inner gradient $dV/dR$ for $R \rightarrow 0$ behaves
precisely as expected for the observed distribution of stellar mass  \cite{Lelli13,Lelli14}.
We thus find ourselves in the odd situation that the dynamics start at small radii behaving exactly as expected 
($V \propto \sqrt{M_*/R}$), but by the time we reach large radii only the total baryonic mass
matters ($V_f \propto M_b^{1/4}$), not its distribution (the dependence on $R$ is lost) \cite{zwaanTF,MdB98a,CR,dBM1996,TVbimodal,noordTF}.  
We explore this quixotic situation in the following sections.

\section{The Global Surface Brightness Dependence of the Mass Discrepancy}
\label{sec:MLSB}

One very basic question we can ask is what the global mass-to-light ratios of galaxies are.  In doing so,
we immediately encounter a problem.  While the total luminosities of galaxies are finite and well-measured,
their masses are not.  The Keplerian decline that should accompany enclosure of the total mass is not detected.

\begin{figure}
\includegraphics[width=17cm]{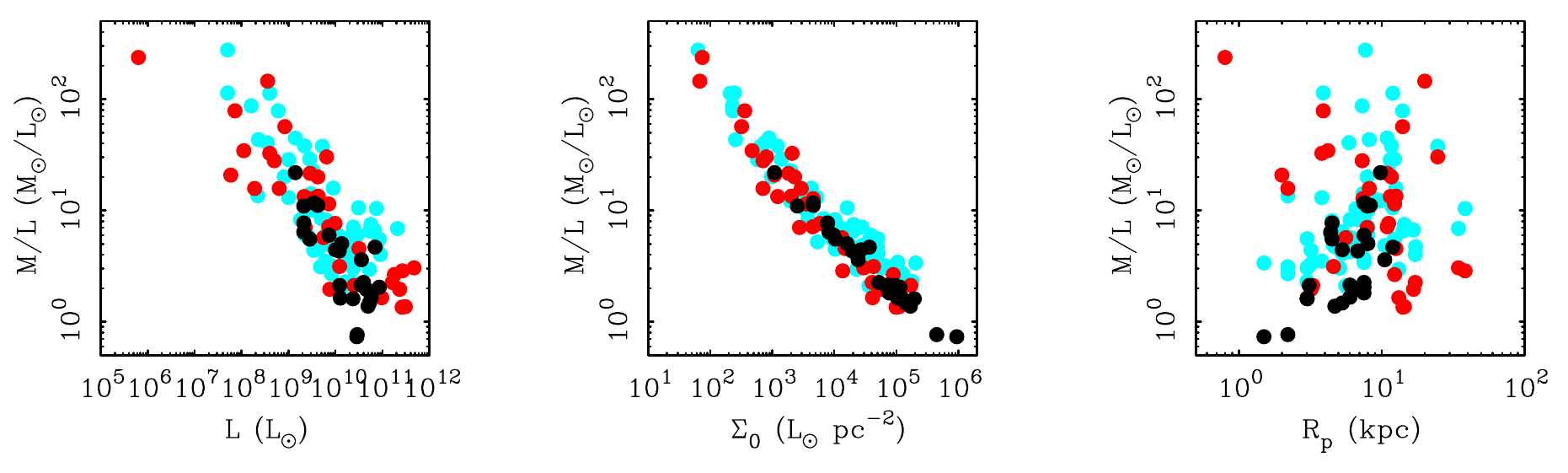}
\caption{The mass-to-light ratio of rotationally supported galaxies enclosed within $2 R_p$.
Each point represents one galaxy.
The choice of radius matters only to the amplitude of $M/L$.  Since $M \propto RV^2$,
the illustrated phenomenon follows for any choice of radius large enough to reach the flat portion of the rotation curve.
The enclosed mass-to-light ratio is shown
as a function of luminosity (left), central surface brightness (center), and size (right) for 
galaxies measured in the $B$-band \cite[blue points]{M05,myPRL}, $K$-band \cite[black points]{SV98,verhTF}, and $3.6 \mu$ \cite[red points]{THINGS,SM2014}.
The strong correlation of the enclosed mass-to-light ratio with surface brightness is a general phenomenon \cite{MdB98a}.
Since $L \propto \Sigma_0 R_p^2$, the weaker correlation with luminosity is simply the convolution of the 
correlation with surface brightness and the lack of correlation with size.}
\label{fig:ML}
\end{figure}

For a spherical mass distribution, the mass enclosed at radius $R$ is simply 

\begin{equation}
M(<R) = RV^2/G.
\label{eqn:sphereM}
\end{equation}

Disk galaxies are not spheres, but this approximation suffices for the present discussion (but not in \S \ref{sec:mdacc}).
The trouble is that we do not know what radius to use in equation \ref{eqn:sphereM}: the extent of the mass distribution
is not limited by the data for any given galaxy.  This remains a problem even for the Milky Way, where inclusion or exclusion
of Leo I as a bound tracer makes a factor of two difference to the total mass.  In general, the maximum radius to which we observe is limited by an arbitrary
combination of observational sensitivity and the intrinsic visibility of whatever tracer we pursue.  No matter how hard
we look, we see $V \approx$ constant.  Sometimes a gradual decline is inferred, but never a Keplerian cutoff.  

In the absence of a clearly defined edge to the mass distribution, the only yardstick available to us is that of the light.
The azimuthally averaged radial light distribution of spiral galaxies is generally approximated as an exponential disk:

\begin{equation}
\Sigma(R) = \Sigma_d e^{-R/R_d},
\label{eqn:expdisk}
\end{equation}

where $\Sigma$ is the surface brightness.  In the exponential disk approximation, $\Sigma_d$ is the central
surface brightness [$\Sigma(R=0)$] and $R_d$ the disk scale length.  This disk scale length provides the most
natural measure of the linear size of the stellar distribution, while $\Sigma_d$ characterizes whether a disk is high surface brightness (HSB)
or low surface brightness (LSB).

While the exponential disk is a tolerable approximation of the starlight in spiral galaxies, it is not perfect.
Many galaxies have a central bulge component that is not fit by equation \ref{eqn:expdisk}.  
NGC 7331 is an example of such a galaxy.
In others there is little bulge, but the mass of gas can exceed that in stars.
DDO 154 is an example of such a galaxy.
Bulges are more concentrated than stellar disks, while the neutral gas is generally more extended.  

In order to incorporate these additional components, \cite{myPRL} defined a mass-equivalent exponential disk (MEED).
The MEED includes all baryonic mass, both bulge and disk stars and gas ($M_b = M_*+M_g$ as in the 
Baryonic Tully-Fisher relation).  It adopts a dynamical measure for the scale length, $R_p$ (Fig.~\ref{fig:RC}).  
The effective central surface mass density of the MEED is

\begin{equation}
\Sigma_b = \frac{3}{4} \frac{M_b}{R_p^2}.
\label{eqn:MEED}
\end{equation}

The factor of 3/4 appears to make this the equivalent of the exponential disk.
For a galaxy with a purely exponential mass distribution, $\Sigma_d$ and $\Sigma_b$ coincide.

The dynamical radius $R_p$ is illustrated in Fig.~\ref{fig:RC}.
It is the radius where the rotation curve of the baryonic mass model peaks.
Since we observe the stars and gas, we can invert the Poisson equation to predict the rotation curve for both the
stellar component $V_*(R)$ and the gaseous component $V_g(R)$.
This can be done analytically for an exponential disk \cite{F70}, for which the resulting peak occurs at $R_p = 2.2 R_d$.
In general, the problem must be solved numerically \cite{BT}.  Publicly available codes to accomplish this are
provided in GIPSY \cite{GIPSY} and DISKFIT \cite{DISKFIT}.  

The limiting uncertainty in estimates of the MEED is the mass-to-light ratio of the stars, $\Upsilon_*$.
For the present discussion, we will bypass this 
issue by ignoring the gas and using the observed luminosity as a proxy for stellar mass.
That is, we measure the characteristic central surface brightness 

\begin{equation}
\Sigma_0 = \frac{3}{4}\frac{L}{R_p^2}.
\label{eqn:LEED}
\end{equation}

This differs slightly from $\Sigma_d$ in that it includes light from the bulge component.

Fig.~\ref{fig:ML} shows the dynamical mass-to-light ratio as a function of the luminosity, characteristic surface brightness, and size
of disk galaxies.  The dynamical mass is computed within $2 R_p$ as 

\begin{equation}
M(< 2R_p) = \frac{2 R_p V_f^2}{G}.
\label{eqn:enclosedmass}
\end{equation}  

The choice of radius is made to encompass most of the luminous mass while falling within the observed range of most rotation curves.  
In a few cases there is a modest extrapolation that assumes the rotation curve remains flat.  
Empirically, this is a good if not perfect approximation.  

Three different samples are utilized.  All have rotation curves from resolved 21 cm data cubes, which provide the most radially extended tracers of the rotation curve.  Each sample has photometry in a different band.  The most data are available for galaxies with $B$-band data \cite{MdB98a,M05}. There is uniform $K$-band data for the Ursa Major sample \cite{SV98,verhTF}, which has the additional benefit that cluster members share the same distance, thus minimizing relative errors induced by distance uncertainties. Finally, there is recent Spitzer Space Telescope photometry at $3.6 \mu$ \cite{SM2014}.  All samples show the same result.  

\begin{figure}
\begin{center}
\includegraphics[width=15cm]{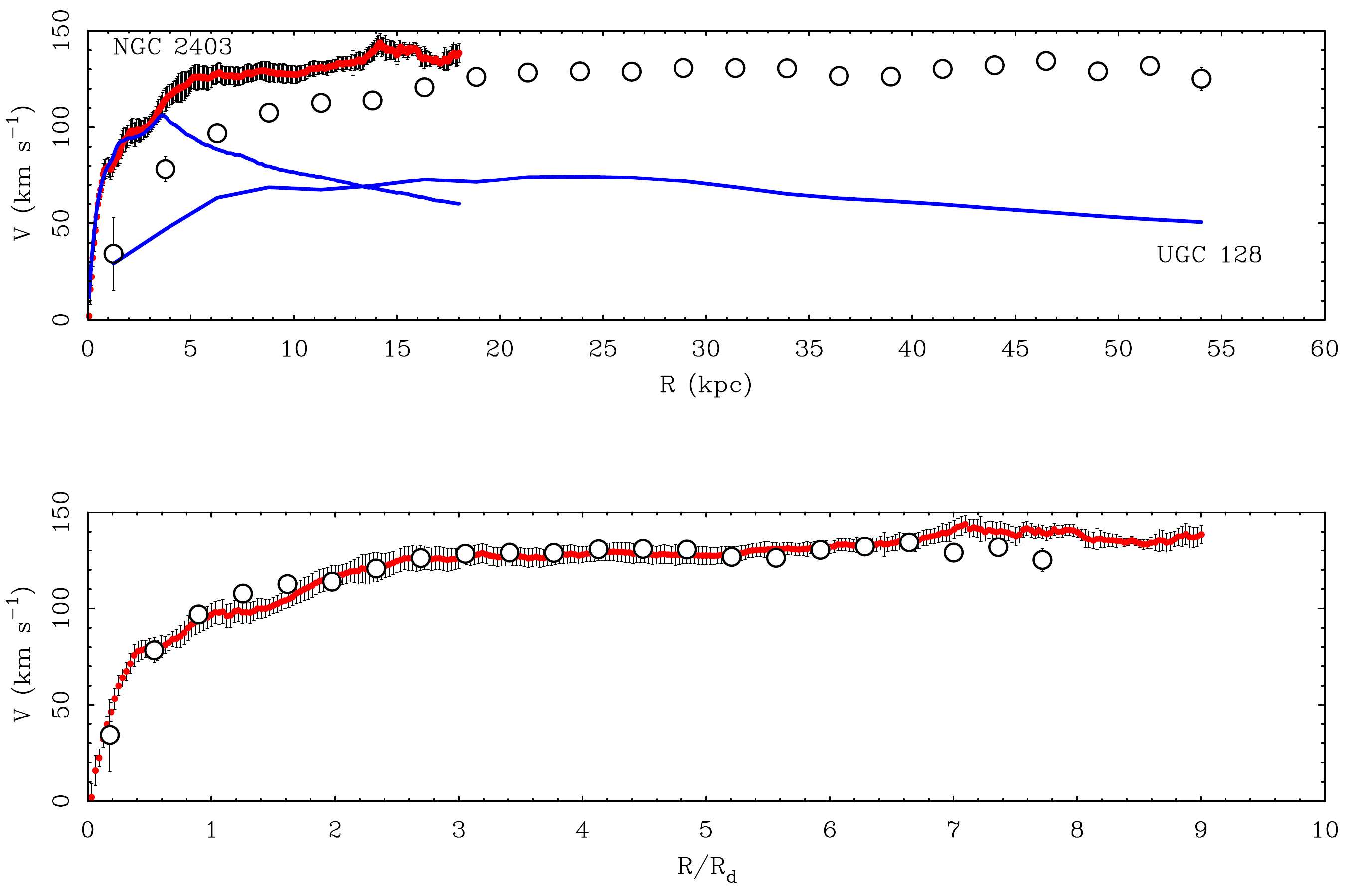}
\end{center}
\caption{The rotation curves of NGC 2403 \cite[red points]{THINGS} and UGC 128 \cite[open circles]{VdB1999} 
plotted against physical radius (top) and again with radius normalized (bottom) to the scale lengths of the optical disks (2 and 7 kpc, respectively).   
Baryonic mass models are also shown in the top panel (lines). The mass model of the more extended LSB galaxy peaks at lower velocity
simply because $V^2 \propto M/R$: the masses are similar but the radii are different.
These galaxies are indistinguishable on the Baryonic Tully-Fisher relation, having very nearly the same ($M_b,V_f$).
Their rotation curves are readily distinguishable in physical units (top panel), but are similar when normalized by the scale length of the disk
(bottom panel).  The dynamics ``knows'' about the distribution of baryonic mass: 
these galaxies not only share a global $V_f(M_b)$; the entire shape of the rotation curve $V(R/R_d | M_b)$ is similar.
}
\label{fig:ngc2403ugc128}
\end{figure}

The enclosed dynamical mass-to-light ratio correlates most strongly with the MEED characteristic surface brightness 
(Fig.~\ref{fig:ML}).  It also correlates with luminosity, but not with size.  Indeed, the correlation with surface brightness
is nearly perfect.  The correlation with luminosity follows simply from the convolution of the good correlation with surface brightness with
the poor one with size \cite{MdB98a} since these quantities are related through equation \ref{eqn:MEED}.
Consequently, surface brightness appears to be more fundamental to the mass discrepancy problem than luminosity.

Measuring the dynamical mass requires choosing a radius within which to measure it.  
The correlation with surface brightness is inevitable for any choice of radius based on the extent of the luminous matter.
This is the only yardstick available.  

One would like to know the extent of the dark matter halo, but this cannot be seen.
One must be careful about including this component, as the enclosed mass one obtains will depend on the
assumption made about the extent of the halo.  For example, it is tempting to assume that galaxies of a given luminosity 
reside in halos of the same total mass.  The true global mass-to-light ratio is thus the same for
galaxies of the same luminosity irrespective of their surface brightness.  This might be true, but it is entirely an
artifact of the assumption.

The key observational result is that the principle dependence of the dynamical mass-to-light ratio is on 
surface brightness \cite{zwaanTF,MdB98a}.  The lower the surface brightness of a galaxy, the larger the mass discrepancy.
HSB galaxies require little dark matter within their optical radius (Fig.~\ref{fig:ML}).
LSB galaxies require quite a bit.  This is not a dichotomy, but rather a continuous progression
with the amplitude of the mass discrepancy increasing steadily as surface brightness declines.

\section{The Local Relation Between Mass and Light}
\label{sec:MLR}

The Tully-Fisher relation and Fig.~\ref{fig:ML} teach us something about the \textit{global} relation between light and mass.
There is also a \textit{local} connection between the two.
That the dynamics ``knows'' about the distribution of light has been known a long time, having been discussed at least as early as 1980 \cite{Rubin1980}.

\subsection{The Universal Rotation Curve}
\label{sec:URC}

One attempt to quantify the relation between the distribution of light and mass was made with the Universal Rotation Curve \cite[URC:][]{URC}.
The URC is posed as a relation between the shape of the rotation curve and the luminosity of a galaxy.  Bright galaxies have steeply rising rotation curves that
decline slowly after a peak.  Low luminosity galaxies have slowly rising rotation curves that only gradually flatten (see Fig.~\ref{fig:RC}).  There is a continuous
variation between these extremes \cite{URC96}.

It is important to realize that luminosity is not the only variable in the URC.  The optical scale length of the disk also enters in this formulation \cite{URC}.
While it is possible to quibble with the detailed formulation of the URC \cite[e.g.,][]{noordURC}, there is a clear consensus that the shape of the
rotation curve depends on the shape of the light distribution \cite{dBM1996,TVbimodal,VdB1999}.  

Galaxies that have the same luminosity have the same flat amplitude of the rotation curve: they are indistinguishable in the Tully-Fisher plane.
However, such identical Tully-Fisher pairs do not share identical rotation curves:  $V(R)$ can differ even if $V_f$ is the same.
High surface brightness galaxies have shorter scale lengths than low surface brightness galaxies of the same luminosity.
Similarly, they have more rapidly rising rotation curves \cite{dBM1996,TVbimodal,Lelli13,Lelli14}: compact light distributions correspond to steep gradients in the 
gravitational potential while diffuse light distributions correspond to shallow potential gradients.  
While $V(R)$ differs when radius is measured in physical units (kpc), normalizing by
the scale length removes most differences (Fig.~\ref{fig:ngc2403ugc128}) so that $V(R/R_d)$ is much harder to distinguish \cite{VdB1999}.  
The dynamics know intimately about the distribution of light.

\subsection{Renzo's Rule}
\label{sec:RR}

The connection between light and mass is also manifest in subtle details.  This is summarized by Renzo's Rule \cite{renzorule}: 
\textit{``For any feature in the luminosity profile there is a corresponding feature in the rotation curve and vice versa.''}

\begin{figure}
\includegraphics[width=17cm]{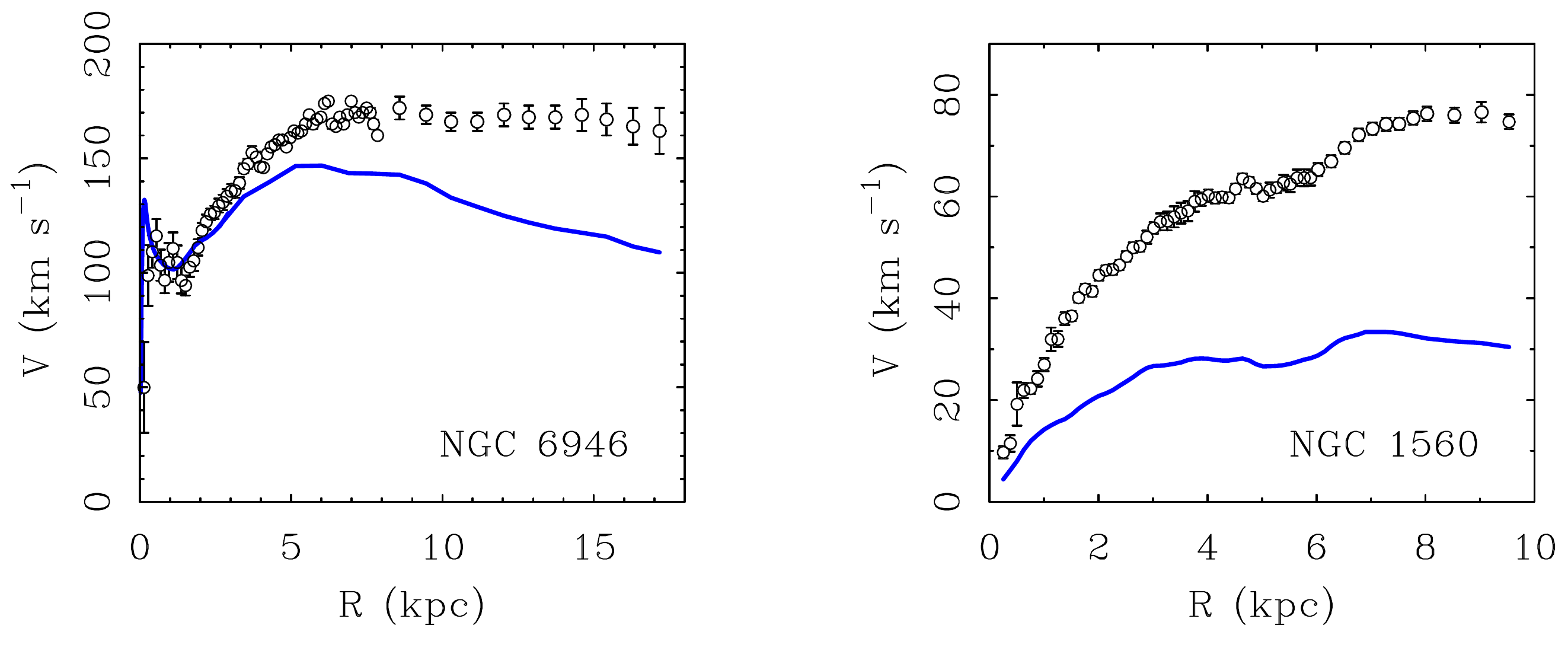}
\caption{The rotation curves of NGC 6946 \cite[left]{boomsma,BOAC,SINGSFP} and NGC 1560 \cite[right]{broeils,gentileN1560} 
together with their baryonic mass models (lines).
These cases illustrate Renzo's rule in both high and low surface brightness galaxies.  In the case of NGC 6946, the tiny, compact 
bulge (just 4\% of the $K$-band light) creates a distinctive feature in the rotation curve at small radii, as expected
from the baryonic mass model.  In NGC 1560, a broad feature in the gas distribution is reflected in the rotation curve.   
Note that unlike the bulge in NGC 6946, the feature in NGC 1560 occurs where the mass discrepancy is large. 
}
\label{fig:ngc6946ngc1560}
\end{figure}

Galaxies with strong central concentrations of stars like a bulge component have steeply rising rotation curves that then fall, 
and may even rise again (Fig.~\ref{fig:ngc6946ngc1560}).  Galaxies lacking bulges lack this double-humped morphology.
In general, the distribution of the dynamics is well predicted by that of the observed luminous matter \cite{Swaters2012}.

One might attribute the correspondence of features in the light distribution with those in the kinematics to maximal disks: the features correspond because
the stars dominate the potential.  This may well be the case in HSB galaxies, but Renzo's rule also applies to LSB galaxies. 
This is surprising, as their dynamical mass-to-light ratios are high down to small radii (Fig.~\ref{fig:MDR}),
implying that their dynamics are dominated by the dark matter halo.
As a dynamically hot, pressure supported, quasi-spherical mass distribution, the dark matter halo cannot support the same features
as the disk \cite{BT}.  Nevertheless, features seen in the baryon distribution are reflected in the kinematics.
For example, the gas in the outer regions of NGC 1560 has a corresponding feature in the total rotation curve (Fig.~\ref{fig:ngc6946ngc1560}).
This is a dramatic example of the general rule in LSB galaxies, where the distribution of the baryons remains predictive of the dynamics
despite their apparent dark matter domination \cite{dBM98,LivRev}.

The effects Renzo's Rule can already be seen in Fig.~\ref{fig:RC}.
The rotation curve of the bright, HSB galaxy NGC 7331 rises steeply, just as predicted by the observed distribution of stars.  
That of the LSB galaxy DDO 154 rises more gradually, and exceeds the expectation from the observed baryonic mass at all radii.  
Nevertheless, there are distinctive features that are shared by both the baryonic mass model and the total rotation curve.
A kink in $V_b(R)$ at $\sim 2$ kpc also appears in the total rotation curve $V(R)$.  Another kink appears at $\sim 5$ kpc.  
The dynamics reflects the distribution of light as described by Renzo's Rule.  

\subsection{The Mass Discrepancy--Acceleration Relation}
\label{sec:mdacc}

The amplitude of the mass discrepancy depends on the surface brightness locally as well as globally.
At each point along a rotation curve, the observed velocity depends on the surface brightness at the corresponding
location.  Thus, while the amplitude of the mass discrepancy grows as one goes farther out, it remains coupled
to the observed luminous matter \cite{SandersMDacc,renzorule,MDacc}.
Indeed, the observed rotation curve $V(R)$ is essentially just a scaled version of the baryonic rotation curve $V_b(R)$ \cite{Swaters2012}.

\begin{figure}
\begin{center}
\includegraphics[width=12cm]{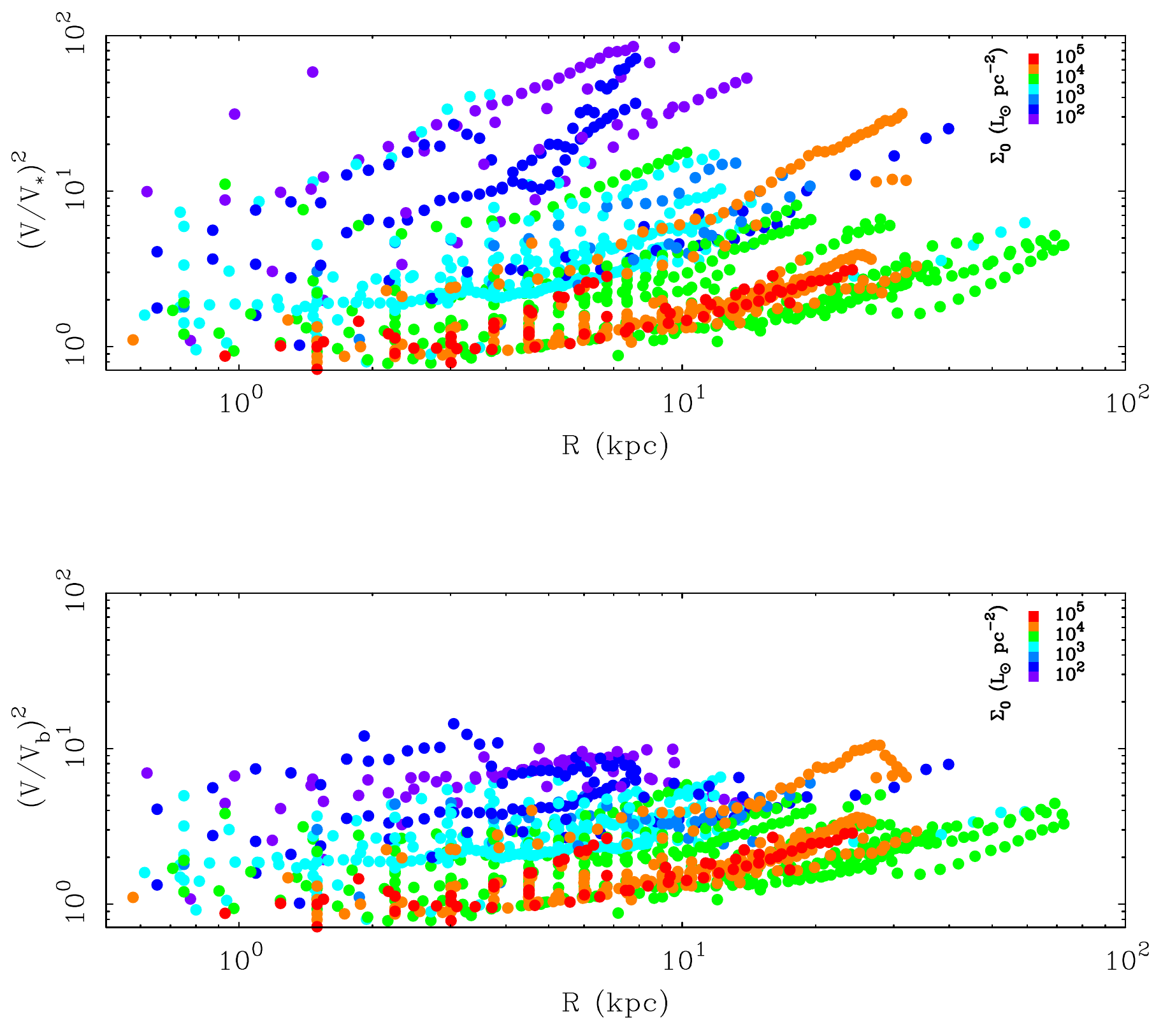}
\end{center}
\caption{The mass discrepancy as a function of radius.  Each point represents one resolved measurement along a rotation curve.
There are 1112 distinct points from 71 galaxies.
At each point, the mass discrepancy $D$ is computed as the squared ratio of the observed velocity to that predicted
by the stars alone ($D = (V/V_*)^2$, top) and stars plus gas ($D = (V/V_b)^2$, bottom).  The mass discrepancy reaches values as large as $D \sim 100$
if only the stars are considered, but this is reduced to $D \sim 10$ when the gas is included. 
In these plots, the stellar mass has been estimated from $B$-band data with mass-to-light ratio estimated from population 
synthesis models \cite{Bell03}.  Individual galaxies are readily distinguishable with progressively
lower surface brightness galaxies having progressively larger mass discrepancies (see inset color code).}
\label{fig:MDR}
\end{figure}

An empirical calibration of the local relation between the mass discrepancy and the luminous mass distribution was offered by \cite{MDacc}.  
The mass discrepancy was defined as the squared ratio of the observed velocity to that predicted by the observed baryons:  

\begin{equation}
D = \frac{V^2}{V_b^2}
\label{eqn:MDacc}
\end{equation}

This ratio is proportional to the ratio of enclosed dynamical to luminous mass.
It uses the full information in the mass model.  There are no simplifying assumptions of sphericity or an exponential disk.

The amplitude of the mass discrepancy does not correlate with radius \cite{SandersMDacc,MDacc}.
While the mass discrepancy grows steadily outwards within an individual galaxy,
the need for dark matter does not appear uniformly as some particular length scale \cite{MdB98a}.
Some large galaxies exhibit little need for dark matter until quite large radii,
while in some small galaxies the amplitude of the mass discrepancy is already large at small radii.
The distinction depends on surface brightness (Fig.~\ref{fig:MDR}).

\begin{figure}
\begin{center}
\includegraphics[width=13cm]{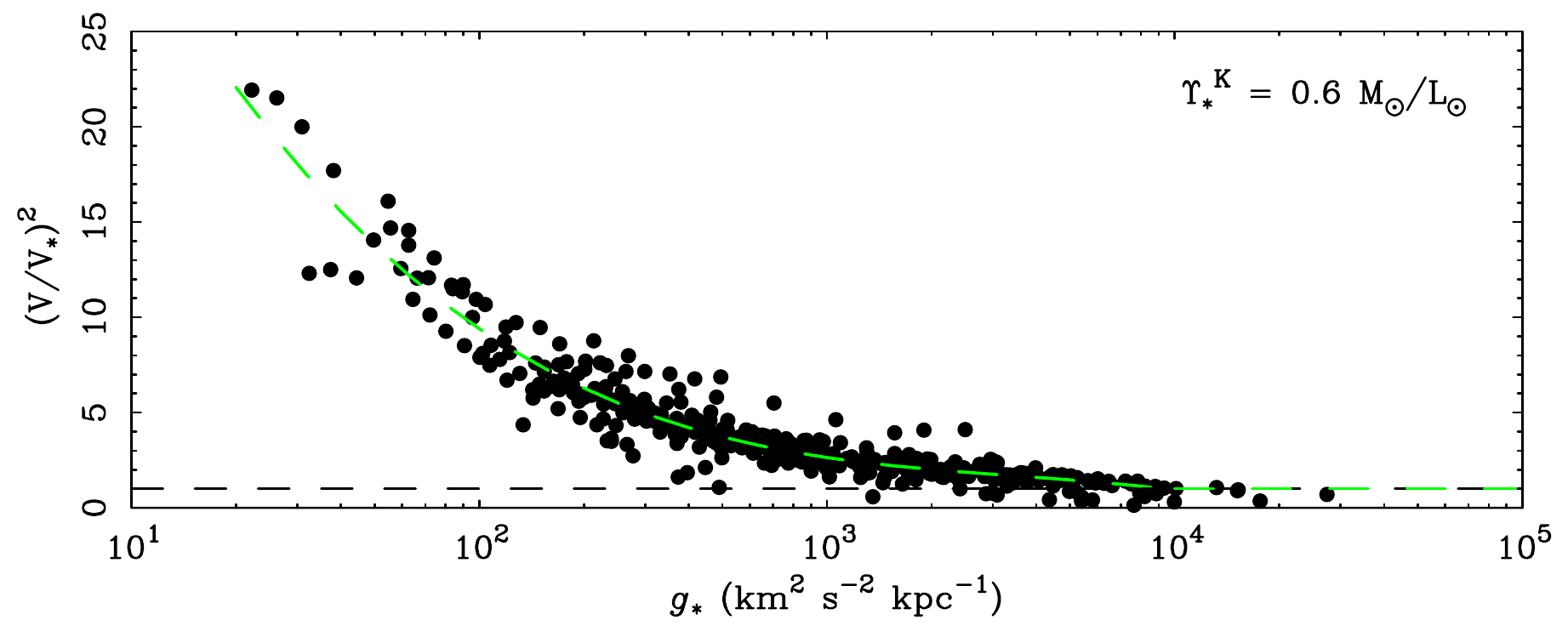}
\end{center}
\caption{The mass discrepancy observed in the Ursa Major sample of 
galaxies \cite{SV98}.  Each point represents one resolved measurement along a rotation curve.
There are 386 distinct points from 30 galaxies, all with $K$-band photometry.
At each point, the mass discrepancy is computed as the squared ratio of the observed velocity to that predicted
by the stars alone: $D_* = (V/V_*)^2$.  
The latter is computed for an assumed stellar mass-to-light ratio $\Upsilon_*^K = 0.6\;\mathrm{M}_{\odot}/\mathrm{L}_{\odot}$
\cite{MS2014}.  The same value is taken for all galaxies so as to directly trace the gravitational potential
predicted by the observed distribution of light.  Individual galaxies do not distinguish themselves
when the mass discrepancy is plotted against the acceleration predicted by the star light, $g_* = - \partial \Phi_*/\partial R = V_*^2/R$.
The observed surface brightness distribution from which $g_*$ is computed is a good predictor of the \textit{total} mass, including dark matter. 
Fitting a polynomial to these data gives the dashed line (eq.~\ref{eqn:KstarsMDacc}).}
\label{fig:MDK}
\end{figure}

The amplitude of the mass discrepancy does correlate with acceleration (Fig.~\ref{fig:MDK}).  
This has been explored in detail by \cite{MDacc}.  For illustration here,
we utilize the Ursa Major sample \cite{SV98}, which has uniform $K$-band photometry for galaxies
at the same distance.  In order to be as empirical as possible,
we compute the velocity expected for the observed stars assuming a constant  
$\Upsilon_*^K = 0.6\;M_{\odot}/L_{\odot}$ for all 30 UMa galaxies.  
This choice of mass-to-light ratio is motivated by recent self-consistent stellar population models \cite{MS2014}.
Fig.~\ref{fig:MDK} for $V_*(R)$ thus shows a direct representation of the observed light distribution.

By construction of the quantities of interest, the discrepancy $D$ for any one galaxy scales inversely with acceleration.
However, there is no reason to expect that all galaxies will follow a single relation between $D$ and $g$.
Indeed, the \textit{a priori} expectation is that they should not \cite{MdB98a,myCJP}.

Nevertheless, the rotation curves for individual galaxies collapse to a single relation between mass discrepancy and 
acceleration.  Individual objects are easily perceptible in Fig.~\ref{fig:MDR}, but not in Fig.~\ref{fig:MDK}.  This behavior persists
for any plausible choice of mass-to-light ratio \cite{MDacc}.  
Again, the surface brightness is playing a fundamental role in the mass discrepancy problem. 

\begin{figure}
\begin{center}
\includegraphics[width=13cm]{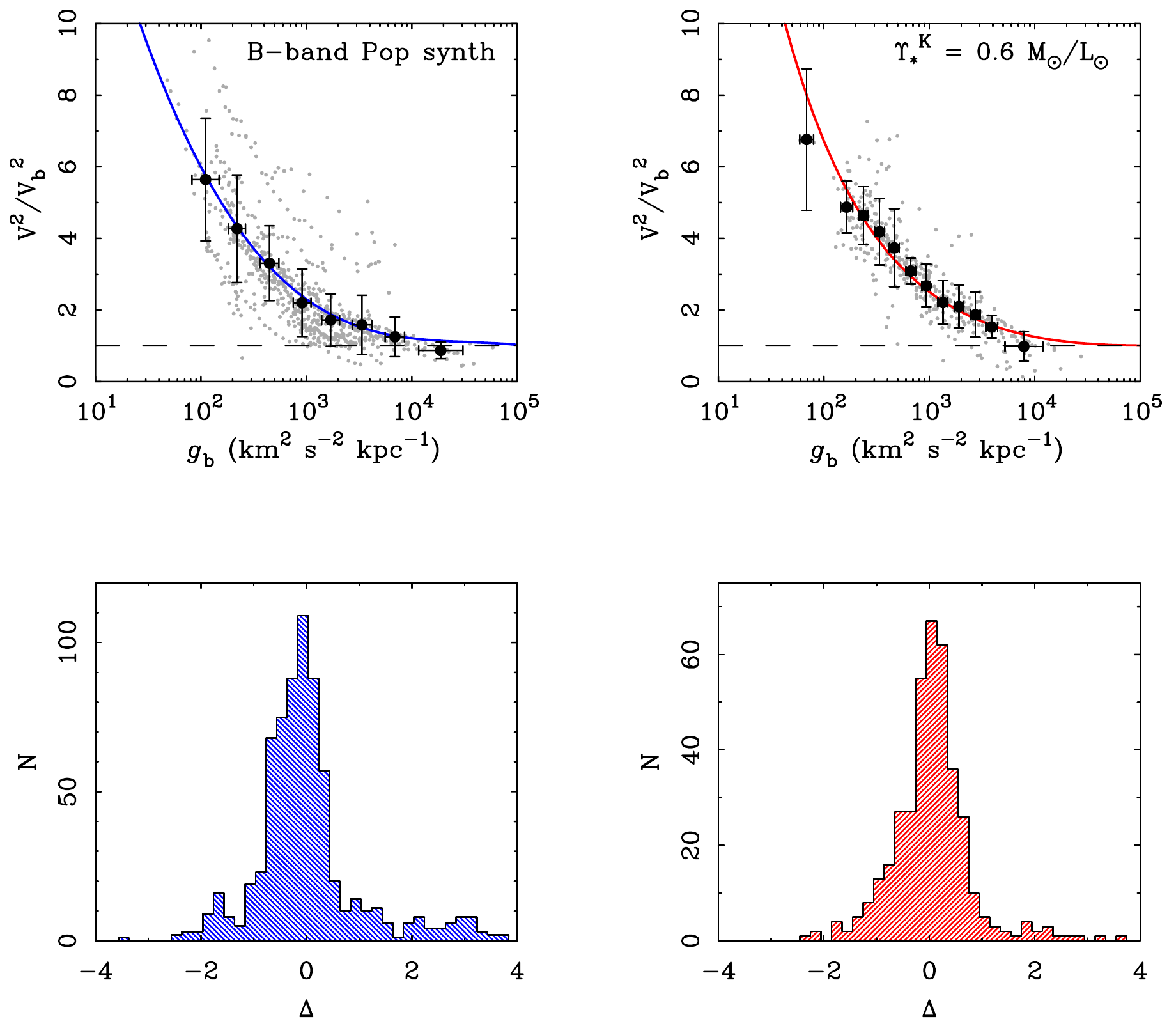}
\end{center}
\caption{The mass discrepancy $D = (V/V_b)^2$ plotted against the gravitational force per unit mass predicted by the baryon 
distribution $g_b = V_b^2/R$.  Left: $B$-band data with mass-to-light ratios predicted by population synthesis 
models \cite{Bell03}. Right: $K$-band data assuming a constant $\Upsilon_*^K = 0.6\;\mathrm{M}_{\odot}/\mathrm{L}_{\odot}$.
Both individual data points (grey) and binned data (dark points) are shown.  Also shown are lines fit to the binned data
(eq.~\ref{eqn:BMDacc}, left, and eq~\ref{eqn:KMDacc}, right). 
Histograms of the residuals around the fits are shown in the lower panels.}
\label{fig:MDcalib}
\end{figure}

The mass discrepancy--acceleration relation \cite{MDacc} illustrated in Fig.~\ref{fig:MDcalib} quantifies Renzo's Rule.  
The ratio between the observed rotation curve and that predicted by the observed distribution of stars and gas correlates with acceleration. 
The acceleration $V_*^2/R$ is computed from $\Sigma_*(R)$ via the Poisson equation.  
The consequence of this empirical relation is that the dynamics can be predicted entirely from the observed distribution of 
stars and gas \cite[a point made early on by][]{SV98}.  Indeed, a near-infrared stellar mass model does well even in the 
absence of observations of the gas distribution (Fig.~\ref{fig:MDK}), at least for star dominated galaxies.
Remarkably, the rotation curve can be predicted from the observed baryon distribution even when the dynamics are 
inferred to be entirely dominated by dark matter.

Quantitative calibration of the mass discrepancy--acceleration relation is discussed by \cite{MDacc}.
The largest uncertainty there was in the estimation of $B$-band mass-to-light ratios.  
The relation persists for any plausible (non-zero) choice of $\Upsilon_*$, but some prescriptions for estimating the mass-to-light ratio are better than others
in terms of the resultant scatter in the relation.

Here I give a brief update.  
Fitting a polynomial to the $K$-band data in Fig.~\ref{fig:MDK} gives a stars-only relation

\begin{equation}
D_* = 75-59.39x+16.37x^2-1.538x^3 \; (\mathrm{K \; band; \; stars\;only})
\label{eqn:KstarsMDacc}
\end{equation}

where $x = \log_{10}(g_*)$ with $g_* = V_*^2/R$ being the centripetal acceleration predicted by the 
observed distribution of stars in Galactic units (km$^2$ s$^{-2}$ kpc$^{-1}$).
At present there exists considerably more $B$-band data for which this calibration can be made.  
Again fitting a polynomial, this time including the gas as well as the stars, we obtain

\begin{equation}
D = 27.9-17.57y+3.85y^2-0.283y^3 \; (\mathrm{B \; band; \; stars \; and \; gas})
\label{eqn:BMDacc}
\end{equation}

where now $y = \log_{10}(g_b)$ and $g_b = V_b^2/R$ is the gravitational force per unit mass from the observed baryons, 
both stars and gas, also in Galactic units.
Note that the amplitude of the discrepancy is reduced by including the gas so that $D < D_*$.
Note also that the rotation curves of a few individual galaxies are perceptible as outliers from the bulk of the
data in the top left panel of Fig.~\ref{fig:MDcalib}.  This is expected with $B$-band data, as the scatter in 
the stellar mass-to-light ratio predicted by population models is large, and one expects to miss the correct value
once in a while.

The variation in stellar mass-to-light ratios is expected to be small in the near-infrared.
Restricting ourselves to $K$-band data and fitting the functional form suggested by \cite{M08},

\begin{equation}
D = (1-e^{-\sqrt{g_b/a_{\dagger}}})^{-1} \; (\mathrm{K \; band; \; stars \; and \; gas})
\label{eqn:KMDacc}
\end{equation}

with best fit parameter $a_{\dagger} = 3800 \pm 100\;\mathrm{km}^{2}\,\mathrm{s}^{-2}\,\mathrm{kpc}^{-1}$.
This functional form imposes well-behaved limits outside the range of the data, and is not subject to the wiggles that can afflict polynomial fits. 
I am not aware of a theoretical motivation to prefer one functional form over others.
Empirically, the functional forms considered here provide adequate descriptions of the data.

\section{Discussion}
\label{sec:interp}

The mass discrepancy--acceleration relation is in some sense a local version of the Tully-Fisher relation.
That there should be a tight connection between light and mass is by no means obvious \cite{MdB98a,myCJP}.  
Indeed, it is one of the most profound observations constraining the missing mass problem.

\subsection{The Dark Matter Distribution}

It is tempting to suspect that the observed coupling between mass and light results from the process of galaxy formation, with
mass telling light where to go: the baryons fall where the dark matter dictates.  This does not at present provide a satisfactory explanation
for the observed phenomenon, as there is no agreed way to predict the distribution of baryons from the dark matter.  While the form of dark matter halos
expected from simulations \cite{NFW,NFWnew} appears to be settled \cite[but see][]{Baushev}, 
the results of baryons forming galaxies within those halos is completely unsettled.

In practice it is the light that tells the mass where to be.
It is possible to predict the distribution of dark matter from the observed distribution of baryons \cite{MDacc,bradamaxacc}.  
Conventionally we write

\begin{equation}
V^2(R) = V_b^2(R)+V_{DM}^2(R)
\end{equation}

where $V_{DM}(R)$ is the circular velocity due to the dark matter halo.  Combining this with eq.~\ref{eqn:MDacc}, we have

\begin{equation}
V_{DM}^2(R) = (D-1) V_b^2(R).
\label{eqn:VDM}
\end{equation}

Note that the right hand side of eq.~\ref{eqn:VDM} depends only on observable quantities, with no reference to dark matter. 
The distribution of baryons completely specifies that of the dark matter, not the other way around.
It is as if the baryonic tail wags the dark matter dog.

There are no free parameters in eq.~\ref{eqn:VDM}.  Once $D(g_b)$ is calibrated, the situation is completely deterministic.
However, the calibration of $D$ depends on the mass-to-light ratio of the stars.  Interestingly, choosing $\Upsilon_*$ to minimize the scatter in $D$
also minimizes the scatter in the Baryonic Tully-Fisher relation \cite{M05}, making it tempting to conclude that this is the `right' mass-to-light ratio.
Indeed, this choice of mass-to-light ratio is in exceedingly good agreement with population synthesis models, matching their normalization, color slope,
and the expected band-dependent scatter \cite{M05,LivRev}.  However, it does allow for one `hidden' degree of freedom, as the mass-to-light ratio of
any one galaxy might not exactly match the one that puts it on the mass discrepancy--acceleration relation. Any deviation of this sort must be small or rare, or
the relation would not persist.

The one free parameter \cite{MDacc} is the ratio of the `true' mass-to-light ratio to that which agrees with the 
mass discrepancy--acceleration relation: 

\begin{equation}
{\cal Q} = \frac{\Upsilon_*^{\mathrm{true}}}{\Upsilon_*^{\mathrm{MDacc}}}.
\end{equation}  

Adjustments to the dark matter halo can be made to compensate in a way that saves
the phenomenology.  Eq.~\ref{eqn:VDM} then becomes more complicated:

\begin{equation}
V_{DM}^2(R) = (\frac{D}{{\cal Q}}-1) V_*^2(R) + (D-1) V_g^2(R)
\label{eqn:generalVDM}
\end{equation}

This is a small additional degree of freedom that affects only the stellar contribution, not the gas.
The shape of the stellar distribution is unchanged; only the amplitude is modulated by ${\cal Q}$.  Since the mass-to-light ratios that inform the mass
discrepancy--acceleration relation are already in good agreement with the expectations of population synthesis, plausible models require ${\cal Q} \approx 1$.
Nevertheless, this one degree of freedom is formally available, so it warrants mention:
eq.~\ref{eqn:generalVDM} is completely general.

The resulting rotation curve of the dark matter halo depends only on the observed baryon distribution. 
It is very nearly universal in form in that the rotation curve of one galaxy's dark matter halo looks very much like that of any others'.
Consequently, one can, to a tolerable if imperfect approximation, write \cite{M07}

\begin{equation}
\frac{V_{DM}}{(\mathrm{km}\,\mathrm{s}^{-1})} \approx 30 \left(\frac{R}{\mathrm{kpc}}\right)^{1/2}. 
\label{eqn:universalhaloV}
\end{equation}

It further follows \cite{walkerandme} that the mass of dark matter enclosed within radius $R$ is 

\begin{equation}
\frac{M_{DM}}{\mathrm{M}_{\odot}} \approx 2 \times 10^8 \left(\frac{R}{\mathrm{kpc}}\right)^2, 
\label{eqn:universalhaloM}
\end{equation}

the density of dark matter varies as

\begin{equation}
\frac{\rho_{DM}}{(\mathrm{M}_{\odot}\,\mathrm{pc}^{-3})} \approx 0.05 \left(\frac{R}{\mathrm{kpc}}\right)^{-1},
\label{eqn:universalhalorho}
\end{equation}

and that the action of dark matter in galaxies is to provide a roughly constant inward acceleration of

\begin{equation}
g_{DM} \approx 900\;\mathrm{km}^2\,\mathrm{s}^{-2}\,\mathrm{kpc}^{-1}
\label{eqn:universalhalog}
\end{equation}

\cite[see also][]{gentile2009,donato2009,walkerandloeb}.

Note that these relations are crude approximations that apply only to the dark matter over a finite range of radii \cite{M07}.
Nevertheless, applying eq.~\ref{eqn:universalhalorho} to the Milky Way yields a local dark matter density of 0.25 GeV cm$^{-3}$,
close to some estimates and within a factor of two of others \cite[e.g,][]{Salucci2010,McMillan2011,bienrave}.

Though obtained from the data for rotating disk galaxies, these relations can have more general utility.  
For example, extrapolating to smaller scales and evaluating eq.~\ref{eqn:universalhaloM} at $R = 0.3$ kpc, 
we obtain $M_{DM} \approx 1.8 \times 10^7\;\mathrm{M}_{\odot}$.
This is consistent with the finding \cite{strigari} that $M(R < 0.3\; \mathrm{kpc}) \sim 10^7\;\mathrm{M}_{\odot}$ in dwarf spheroidals.
Indeed, the above relations \cite{MDacc,M07} were known prior to the work of \cite{strigari}, so the constancy of dark mass at a
fixed radius should not have been surprising.
This particular case highlights the danger of quoting masses involving the unseen dark matter, as it is difficult
to avoid artifacts imposed by the choice of radius within which this is measured \cite[contrast][]{zwaanTF,sprayTF}.

Though useful, one should be careful not to over-interpret these approximate relations.  They do not hold absolutely, and can be rather crude.
Equations \ref{eqn:universalhaloV} -- \ref{eqn:universalhalog} only approximate eq.~\ref{eqn:VDM}, which itself is derived from
the more fundamental mass discrepancy--acceleration relation (eq.~\ref{eqn:MDacc}).

\subsection{Neither Cusps nor Cores}
 
The dark matter distribution in any given galaxy is specified by observations with very little freedom. 
The only free parameter is ${\cal Q}$ in eq.~\ref{eqn:generalVDM}.
To build a conventional disk plus dark matter halo model, one requires at least three parameters: the mass-to-light ratio of the disk and at least two more 
to describe the dark matter halo (typically a scale radius and a velocity or mass scale for some assumed density profile).   This is two more parameters than the
data require.  A great deal of parameter degeneracy is thus expected (and found) when one uses more than the minimum number of parameters necessary
to describe the data.

Traditionally, dark matter halos were assumed to be of the pseudo-isothermal form \cite[e.g.,][]{dBM97} with a constant density core ($\rho_{DM} \propto$ constant).
This provides good fits to the data, but such a model was not supported by subsequent numerical simulations of structure formation \cite{NFW}. 
These show an inner cusp, with a density profile continuously increasing like $\rho_{DM} \propto R^{-1}$ as $R \rightarrow 0$.
This functional form \cite[the NFW halo,][]{NFW} generally provides less satisfactory fits to the data than dark matter halos with constant density cores \cite{dB2010}.  
NFW halos can sometimes, though not always, be made to fit HSB galaxies \cite{THINGS,HarleyNFW}.  
However, they persistently fail to explain the data for LSB galaxies
\cite{OhThings,KdN08,KdN09,KdN11,AdamsSimon2014}. This discrepancy is the `cusp-core' problem.

Considerable effort has been expended trying to solve this cusp-core problem. The problem is typically framed in terms of
the inner slope of the dark matter profile, with it being considered desirable if a model can produce a near-constant density core.
This in itself is not adequate; one must also recover the mass discrepancy--acceleration relation.  
The slope of the dark matter halo profile is not necessarily a cusp or a core; it is whatever the baryon distribution dictates.
It need not be described as a power law at all.

Near the centers of HSB galaxies, the acceleration is large and $D \rightarrow 1$ so there is no clear need for dark matter.
While we reject hollow dark matter halos as unphysical, the data are consistent with such a situation.
Cored dark matter halos generally provide better fits than cusped halos because they can
accommodate $D \approx 1$ by fitting a low core density that contributes negligibly to the rotation at small radii.
However, the situation is often degenerate: in galaxies with bulge components, the stellar distribution is itself cuspy,
so it can be difficult to distinguish between stellar mass and dark mass.

The debate between cusps and cores misses a more fundamental point.
The real challenge is to understand the mass discrepancy--acceleration relation.
If a model can provide a good explanation for eq.~\ref{eqn:MDacc}, than the rest follows.
To address this requires a detailed prediction for the distribution of baryons as well as dark matter.
The only paper of which I am aware that attempts to address this issue is \cite{vdBD2000}, who
simply assume smooth exponential disks.  By construction, this does nothing to address Renzo's rule.

\section{Broader Context}

The National Academy of Sciences \cite{NatAcad} define a Natural Law as
\begin{quote}
A descriptive generalization about how some aspect of the natural world behaves under stated circumstances.
\end{quote}
Renzo's rule (\S \ref{sec:RR}; \cite{renzorule}) is an example of such a descriptive generalization.

I summarize the properties of rotationally supported galaxies in a similar spirit: 
\begin{enumerate}
\item Rotation curves attain an approximately constant velocity that persists indefinitely (flat rotation curves).
\item The observed (baryonic) mass scales as the fourth power of the amplitude of the flat rotation (the Baryonic Tully-Fisher Relation).
\item There is a one-to-one correspondence between the radial force and the observed distribution of baryonic matter
(the mass discrepancy--acceleration relation).
\end{enumerate}
Whether these should be regarded as mere descriptive generalizations or true Laws of Nature is a matter 
for contemplation.  

One important issue is the intrinsic scatter in the second and third law.
I have spent a considerable portion of my career investigating galaxies whose physical properties suggest that they should deviate
from these relations.  So far, they do not.  The scatter in both the Baryonic Tully-Fisher Relation \cite{M11,M12} and the
mass discrepancy--acceleration relation \cite{MDacc} is consistent with that caused by observational uncertainty and the
scatter expected in stellar mass-to-light ratios.  There is precious little room left for intrinsic scatter.
If this holds, then the second and third laws appear more as Natural Laws than happenstance.

I have intentionally written the Laws of Galactic Rotation in analogy with Kepler's Laws of Planetary Motion. 
I do this in part because I suspect they deserve an equivalent status, and in part to distinguish between
data and theory.  Both sets of laws are empirical.  
They demand an explanation regardless of what label we place on them.

Kepler's Laws were discovered empirically, and subsequently explained as a consequence of angular momentum conservation
and Newton's Universal Law of Gravity.  In similar fashion, the three laws of galactic rotation can be explained as a consequence
of MOND \cite{milgrom83}. To paraphrase Newton,
\begin{quote}
\textit{Everything happens in rotating galaxies as if the effective radial force law is MOND.}
\end{quote}

One of the greatly satisfying aspects of Newtonian gravity is its clear explanation for Kepler's Laws. 
MOND plays the same role for the laws of galactic rotation.  Unlike the historical example, important aspects of the data
for galaxies were predicted \textit{a priori} by MOND.  Like dark matter, MOND was motivated by the first law.  
But MOND uniquely predicts that the relation between the amplitude of the flat velocity and mass is absolute.
There should be no intrinsic scatter, nor any residual dependence on the size of 
an object.  Moreover, the slope of the Baryonic Tully-Fisher relation need not now agree with MOND's long standing
prediction, yet the two are consistent. Indeed, MOND can be used to predict \cite{M11} the rotation velocities of galaxies much
smaller than were known when the Tully-Fisher relation was discovered \cite{TForig}.

The third law was entirely anticipated by MOND.  It was not obviously presented in the data at the time MOND was formulated.
No such relation was anticipated in the context of dark matter.

Note the use of \textit{as if} in the paraphrase of Newton.  These words were used by Newton in describing the inverse square law.
No doubt this was in part because he was exceedingly careful in how he phrased important statements.  But it is also worth remembering that 
there was no understanding at the time how one could have action at a distance.  To suppose as much caused great consternation, 
and may have seemed to some almost like magical thinking.

In retrospect, we have a very good understanding of the geometric origins of the inverse square law.  That did not come about overnight.
Hence the use of \textit{as if} remains appropriate.  I have no idea why, physically, MOND is the effective radial force law in galaxies.
Perhaps this will be clear in retrospect at some point in the future.  At present, one can imagine three broad categories of possibilities:
\begin{enumerate}
\item MOND represents a true modification of dynamical laws.
\item The laws of galactic rotation are a consequence of some process during galaxy formation.
\item The properties of dark matter particles impose the observed phenomena.
\end{enumerate}

The first of these possibilities provides the most natural explanation of the three laws.
MOND has had a great deal of predictive success as a theory of dynamics in the limit of low 
acceleration \cite{MdB98b,dBM98,SMmond,M11,LivRev}.  This success is not restricted to rotating systems \cite{MM13a,MM13b,milgrom2014}.
It would be unique in the history of science for a theory with so many quantitatively accurate predictive successes to be utterly devoid of meaning.

The second possibility seems to be most popular at present.
This is due in part to the more widely known successes of $\Lambda$CDM as a description of cosmology.
Unfortunately, where one paradigm is eloquent and predictive, the other tends to be rather mute \cite{myCJP},
making them difficult to compare.  However, the general presumption seems to be that $\Lambda$CDM must be right,
so the issues with galaxies will surely get sorted out.  While comforting, this makes no more sense than saying MOND
must be right, and its issues with cosmology will surely get sorted out.

The third possibility is a hybrid solution that I find aesthetically displeasing. 
However, in the absence of a positive detection of any of the more commonly considered dark matter candidates,
it is worth considering whether it is possible to hypothesize a dark matter candidate that satisfies the three laws of
galactic rotation while simultaneously preserving the successes of $\Lambda$CDM on large scales.
One candidate for this is dipolar dark matter \cite{blanchet,Blanchetnatural}.
More generally, it seems appropriate to consider the three laws as an astrophysical constraint on the nature of dark matter,
and explore what theories may be possible \cite[e.g.,][]{Sandersboson,Khoury}.

I do not find any of the three possible interpretations completely satisfactory.
There are observations that are not fully understood within the framework of any one paradigm, be it $\Lambda$CDM or MOND.
Perhaps that should be seen as a good thing:  there remains fundamental physics yet to be discovered.

\acknowledgements{Acknowledgements}

I am grateful to many colleagues who have worked on rotation curves over the years.
Most especially, I would like to thank Bob Sanders, Thijs van der Hulst, Renzo Sancisi, 
Erwin de Blok, Marc Verheijen, Filippo Fraternali, Federico Lelli, and Vera Rubin
for their insight, hard work, and dedication to astronomy as an empirical science.
This work was supported in part by NASA ADAP grant NNX13AH32G.
This publication was made possible through the support of the John Templeton Foundation. 
The opinions expressed in this are those of the author and do not necessary reflect the views of the John Templeton Foundation.

\authorcontributions{Author Contributions}

This document was composed by the author in its entirety.

\conflictofinterests{Conflicts of Interest}

The author declares no conflict of interest. 

\bibliography{review}
\bibliographystyle{mdpi}

\end{document}